# Managing Level of Detail Through Peripheral Degradation:
# Effects on Search Performance with a Head-Mounted Display

Benjamin Watson

Department of Computing Science, University of Alberta

615 General Services Building, Edmonton, Alberta, Canada T6G 2H1

Tel: +1 403 492 9918; Fax: +1 403 492 1071

Email: watsonb@cs.ualberta.ca

Neff Walker, Larry F. Hodges, & Aileen Worden

Graphics, Visualization & Usability Center, Georgia Inst. Technology

801 Atlantic Drive, Atlanta, GA 30332-0280, USA

Running Head: Peripheral Detail Degradation and Performance with a Head-Mounted Display

**ABSTRACT**

Two user studies were performed to evaluate the effect of level of detail (LOD) degradation in the periphery of head-mounted displays on visual search performance. In the first study, spatial detail was degraded by reducing resolution. In the second study, detail was degraded in the color domain by using grayscale in the periphery. In each study, ten subjects were given a complex search task that required users to indicate whether or not a target object was present among distracters. Subjects used several different displays varying in the amount of detail presented. Frame rate, object location, subject input method, and order of display use were all controlled. The primary dependent measures were search time on correctly performed trials, and the percentage of all trials correctly performed. Results indicated that peripheral LOD degradation can be used to reduce color or spatial visual complexity by almost half in some search tasks without significantly reducing performance.

# Managing Level of Detail through Peripheral Degradation: Effects on Search Performance in a Virtual Environment

## 1. INTRODUCTION

As virtual environments (VE) researchers attempt to broaden the range of applications for VE technology, they are attempting to display ever larger and more complex models. Many of these models, however, cannot be displayed with acceptable frame rates in current systems. Several researchers have identified this *frame* or *update rate* problem as one of the most pressing problems facing the VE community [NSF, 1992; Van Dam, 1993]. Foremost among the proposed solutions to this problem is varying *level of detail* (LOD). As used by most VE researchers, this phrase refers to model and rendering complexity, which can be varied to ensure that VEs are rendered at some minimal acceptable frame rate.

There are many techniques that might be used to generate images of varying detail or complexity, including using geometric models of varying degrees of accuracy [DeRose & Lounsberry, 1993; Rossignac & Borrel, 1992; Turk, 1992; Varshney, Agarwal, Brooks, Wright, & Weber, 1995], lighting and shading models of differing levels of realism, and textures and graphics windows of differing resolution [Maciel & Shirley, 1995; Schaufler, 1996]. Many researchers have compared the relative importance of different graphics rendering techniques in traditional display environments [Atherton & Caporeal, 1985; Barfield, Sandford, & Foley, 1988; Booth, Bryden, Cowan, Morgan, & Plante, 1987]. In general, these studies showed significant effects on performance when image complexity is varied. However, in most cases a point of diminishing returns was reached, beyond which additional image complexity and computation produced little or no increase in user performance. This suggests that varying LOD by using different rendering techniques may be a promising approach to solving the frame rate problem. However, we are not aware of any studies that address the perceptual cost of LOD generation by using models of varying accuracy.

One must not forget that varying detail, like varying frame rates, affects user performance. Performance can be thought of as a function of function of detail and frame rate. Ideally, the combination of frame rate and LOD that maximizes performance should be dynamically located and maintained. We call this task LOD management, and it is very complex.

Fortunately, there is a fairly obvious first step. The human perceptual system does not perceive all the detail in its environment. Thus there will always be a disparity between displayed and perceived detail. Since human performance depends only on perceived detail, this disparity represents wasted computation. By reducing this disparity frame rates can be improved while perceived detail remains unchanged, resulting in improvements in performance.

Several methods of reducing this disparity have already been proposed, and two are already in widespread use. As the size of an object decreases, so does the eye's ability to resolve its detail. Flight simulators and VE systems [Funkhouser & Sequin; 1993, Maciel & Shirley, 1995] exploit this fact by using lower LOD when the visual angle of a portion of the model is small. This technique has a proven track record in the flight simulator industry. The eye's ability to resolve detail also decreases with retinal eccentricity [Bishop, 1986]. This suggests the possibility of a computationally and perceptually efficient divided display containing a central, high detail inset, corresponding to the perceptual characteristics of the foveal area of the retina; as well as a surrounding, simpler periphery, corresponding to the perceptual characteristics of the peripheral area of the retina [Reddy, 1995]. Funkhouser and Sequin, Maciel and Shirley as well as Ohshima, Yamamoto and Tamura [1996] have implemented systems that degrade peripheral detail to improve frame rate. Howlett [1992] and Yoshida, Rolland and Reif [1995] have presented designs for HMDs containing high resolution central insets. Extremely expensive high-end Fiber-Optic HMDs [Barrette, 1986; Fernie, 1995] produced by CAE Electronics include eye-tracked high-detail insets, generally subtending about 25 horizontal degrees. However, we are not aware of any studies evaluating the impact on user performance of managing LOD by degrading peripheral detail.

This paper describes two user studies evaluating the effectiveness of LOD management through peripheral detail degradation in head-mounted displays (HMDs). The studies are similar in design to the study described in [Pausch, Schackelford & Proffitt, 1993]. Because the focus of these studies is LOD management, and not LOD generation, we chose to generate LOD simply by varying display resolution in the first study, and color content in the second. Moreover, because currently available eye tracking technology is unwieldy and expensive, we worked under the assumption that head-tracking alone would allow effective peripheral degradation in an HMD.

Our hypothesis in this study was that peripheral detail degradation would result in minimal perceptual loss and significant computational gain. The computational portion of this assertion has been partially examined in [Funkhouser & Sequin, 1993; Maciel & Shirley, 1995]. We attempted to verify only the perceptual portion of this assertion by measuring subject performance time and accuracy while peripheral detail was degraded over various visual extents and with various LOD.

## 2. EXPERIMENT 1

In our first experiment, we varied LOD by controlling display resolution in the periphery. We used two different peripheral resolutions, and three different high detail inset sizes. We compared these six displays to two insetless displays using low detail resolutions, as well as one undegraded display using high detail resolution. We asked subjects to perform a search task that required heavy use of peripheral display areas. Because visual acuity and sensitivity decrease with eccentricity, we expected that loss of detail would have less impact on subject performance time and accuracy when only peripheral detail was degraded than when detail was degraded across the entire display.

We anticipated that use of the undegraded display would result in the lowest subject performance times and highest accuracies. In addition, we predicted significant differences among the displays with insets, indicating that speed and accuracy declined as LOD and the size of the inset decreased.

A preliminary study [Watson, Walker, & Hodges, 1995] indicated that the utility of peripheral LOD management would be greatly affected by the nature of the user's task. As task difficulty increases, subjects should require greater amounts of visual detail. To test this hypothesis, we varied the difficulty of the search task by varying the number and grouping of objects.

## 2.1 Experimental Methodology

2.1.1 *Participants*. The participants in the experiment were 10 graduate students at the Georgia Institute of Technology. All of the subjects reported 20/40 corrected vision or better (still a significantly higher acuity than the best 20/355 acuity offered by our display, see below).

Because of our emphasis on peripheral visual search, subjects with glasses were not permitted to participate in the experiment.

2.1.2 *Design*. The study utilized a six factor, mixed design. The primary independent variables were all within-subjects variables. These primary independent variables were display (varying in inset size and resolution, see Table 1), number of objects (1, 3, or 5), quadrant (location of the target object in the search space), and grouping (objects in the same or different quadrants). In addition to these primary independent variables (used in the data analyses) the study also had three control variables. These variables were trial condition (target object was either present or absent), number of trials (there were four trials of each of the conditions), and button assignment (either the thumb or the index finger was used to signal target present; this variable was introduced to control possible biases to a certain button configuration). The first two of these control variables were within-subjects, while button assignment was a between-subjects variable. The insetless, coarse resolution (coarse-0) display was used only as a control condition. The performance in this condition was analyzed separately.

2.1.3 *Apparatus*. The images used in this study were generated in real time by a Silicon Graphics Onyx Reality Engine II. We used the gl graphics library and the SVE virtual environments library [Verlinden, Kessler & Hodges, 1993] in our programming. Silicon Graphics' scan converting hardware and software converted these images into an NTSC signal. The scan conversion process requires an input image of 640 pixels in width and 480 pixels in height.

________________

Figure 1 about here

_______________

LOD was varied by changing resolution. When the same LOD was used throughout the display, a single image was generated with the required number of pixels, and then textured onto a 2D polygon with the constant screen size required by our scan converter. When high detail insets used, two images were generated with the needed numbers of pixels, and then textured onto two 2D polygons: one for the low LOD periphery, and one for the high LOD inset, both fitting inside the required constant screen size. The two polygons were overlapped slightly and blended with alpha transparency to make the boundary of the high detail inset harder to detect (see Figure 1). Texturing was accomplished in real time with the fbsubtexload command and FAST_DEFINE [Silicon Graphics, 1993]. Since eye tracking was not used, insets were always located in the center of the displayed image. Insets were rectangular, with the same aspect ratio as the display (see Figure 2).

_______________

Figure 2 about here

_______________

Subjects wore the widely used Virtual Research Flight Helmet to immerse themselves in the experimental environment. The Flight Helmet mounts two color LCD displays on the user's head, each with vertical field of view of 58.4 degrees, and a horizontal FOV of 75.3 degrees. Each LCD contains an array of 208 x 139 color triads (the equivalent of pixels on CRT displays). We used the Flight Helmet in a biocular mode by sending the same image to each of the video inputs, and mounting adhesive plastic fresnel lenses provided by Virtual Research on the last elements of the HMD optics (directly before the eyes). These simple fresnel lenses create appropriate disparity and enable fusion by translating the image along the horizontal interocular axis. When the Helmet is used in a binocular mode, the disparities already in the stereo signal pair make the fresnel lenses unnecessary. The motion of a subject's head in the Flight Helmet was tracked with the Polhemus Isotrak II 3D tracking hardware. Subjects were free to turn and move their heads. All of this motion was represented in the virtual environment.

The LEEP optics in the Flight Helmet introduce a mild nonlinear distortion into the displayed image, so mild that most users of the Helmet are not aware of it unless told it is there. Since the degree of the distortion is a function of the distance from the LEEP optical axis, the edges of the display are most distorted, while the center of the image remains largely undistorted. Some of these authors [Watson & Hodges, 1994] and Robinett and Rolland [1992] have provided a complete description of the LEEP optics and the accompanying distortion, and suggested a method of correcting for it. In practice however, this correction causes too severe a reduction in frame rate, and the distortion remained uncorrected in this study. Because of the distortion, resolution is highest at the center of the Flight Helmet's display, and lowest in the periphery. Average horizontal angular resolution is 21.74 arcmins (which equates to a Snellen fraction of 20/435), with resolution at the center of the display approximately 17.76 arcmins (20/355), and at the edge of the display 23.46 arcmins (20/469). For a complete description of LEEP distortion's relation to resolution, see [Bryson, 1993].

___________________

Figure 3 about here

___________________

Figure 3 summarizes the image generation process for insetted displays. Two versions of the current virtual environment view, one in high detail and one in low detail, are generated at appropriate resolutions on the Onyx screen. These two images are then composited by the texture mapping process into a single final workstation image of constant size. This image is then scanned by SGI hardware and software into an NTSC signal that is sent to the HMD, which filters the signal and displays the resulting image on its 208 x 139 triad screen. This displayed image is then magnified, slightly distorted and translated by the Flight Helmet's LEEP brand and simple fresnel optics. It is important to note that since the number of triads in the Flight Helmet is much smaller than the number of pixels in the constant sized source window on the SGI, the resolution of the final image is limited by the HMD. Note also that the inset is fixed relative to the HMD screen while the virtual environment is not; the net effect is somewhat like looking through a fogged glass mask that has been wiped clean in its center.

Subjects used a plastic mouse shaped like a pistol grip to respond to the experimental environment. The mouse had two buttons for the thumb mounted on top, and one button for the index finger mounted on the front. The mouse was not tracked nor displayed in the virtual environment. When using the experimental environment, subjects stood inside a 4x4 platform raised six inches and surrounded by a three foot railing.

_______________

Figure 4 about here

_______________

The virtual experimental environment consisted of a floor, indicated by a grid of white lines on black. The background above the floor was also black. A home object indicated starting position for search.  It was a rectangular shape extending from the floor and slightly taller than subjects, and was textured with a bullseye design.  Subjects could not see any representation of themselves in the virtual environment (see Figure 4).

2.1.4 *Procedures*.  Each experimental trial consisted of a single search task.  After focusing on the home object, subjects pressed the right thumb button to begin the task.  After a random (between .1 and .8 seconds) delay, the home object disappeared, and objects appeared to the right of the subject's initial view.  Subjects attempted to locate a unique target object and pressed one of two buttons to indicate the presence or absence of the target object.  Each subject used the same button configuration throughout the experiment.  The objects then disappeared, and the home object reappeared, with on screen feedback indicating correctness and time of response. When the subjects had again focused on the home object and pressed the appropriate button, a new search task began.  On target-present trials, subjects were not credited with a correct trial unless they had brought the target object onto the HMD screen.  This eliminated simple guessing as a search strategy.  In trials without a target object, subjects were not credited with a correct trial unless they had brought every object onto the HMD screen.  This forced an exhaustive search.

Subjects performed the search task with each of nine display types (see again Table 1). Resolution was varied at three levels by varying the number of pixels in the initial image (or images).  At the fine level of display resolution, the image scanned into the HMD was 25% of NTSC: 320 x 240 pixels (14.12 horizontal arcmins/pixel over an FOV of 75.3).  This resolution is greater than the Flight Helmet's resolution, and ensured the HMD resolution was at its best. Medium resolution was 9%  of NTSC: 192 x 144 (23.62 arcmins).  This resolution is slightly lower than HMD resolution.  Coarse resolution was only 1% of NTSC: 64 x 48 (70.59 arcmins), much less than HMD resolution and making discrimination of crucial object features difficult. The high detail inset was always presented at the fine level of resolution. Because display on the HMD screen is performed before the optics introduce distortion, eccentricity-dependent variations in angular resolution occur in equal magnitude at all three resolution levels (roughly 18 to 23 at fine resolution, 21 to 26 at medium resolution, and 68 to 73 at coarse resolution).

_______________________

Figure 5 about here

_______________________

The amount of display area degraded was controlled by varying the size of the high detail inset. The inset size was varied at five levels: all of display area (undegraded), 25% of display area (half of the display's height and width; 37.65 horizontal degrees), 9% (30% of height and width; 22.59 degrees), 2.25% (15% of height and width; 11.29 degrees), and 0% (insetless) (see Figure 5). The size of the image generated for texturing into the high detail inset was adjusted to ensure constant pixel size corresponding to the fine level of resolution.

The column of Table 1 labeled "Total Pixels in Wkstn" shows the relative amounts of detail contained in the displays as measured by pixels in the final workstation image. However, because the HMD screen had far fewer pixels than the workstation image, a more accurate measure of detail are the counts of pixels in the HMD for each display, shown in the column of Table 1 labeled "Approx Pixels in HMD". Because the reduction of the high resolution NTSC signal to the lower resolution HMD screen was done in the Flight Helmet hardware, we cannot know the exact counts. However, we have approximated these counts[1] and presented them in Figure 6. The figure shows a gradual decline of pixels for the displays with a medium resolution periphery, and a much steeper decline for displays with a coarse resolution periphery.

______________________

Figure 6 about here

______________________

During each trial, subjects searched through several randomly located, identically sized objects for a target object. Most objects were textured with identical images of a smiling human face. The single target object was textured with the same face, with the mouth closed (see again Figure 2). Objects always appeared at the same virtual distance, and subtended a horizontal visual angle of approximately 12 degrees. The mouth on the face subtended a visual angle of approximately 2 degrees horizontally and 1 degree vertically. Since the smallest inset subtended 11.29 horizontal degrees, subjects were always able to place almost an entire object into any high detail inset. The mouth on any object's face fit quite easily within all insets. Because the inset was fixed relative to the HMD and the objects in the virtual environment were not, all objects could potentially lie entirely within the low detail periphery, entirely within the high detail inset, or on the boundary between them, depending on the subject's current head position.

Object location was controlled through the use of four contiguous quadrants located in a two by two grid around the subject (see Figures 4 and 7). These quadrants were of equal area, and none were located above or below the subject.  No quadrants were in view when the subject was in home position (looking at the home object).  The horizontal angular extent of each quadrant was 75.3 degrees (equaling the HMD's horizontal field of view), the vertical extent was 49.3 degrees.  Each quadrant fit entirely into a single view.  The vertical boundary between the quadrants was roughly at eye level.  Our pilot study, in which quadrants were evenly distributed around the subject, showed no interactions of quadrant with display.  This enabled us to reduce average search time by placing the quadrants only behind and to the right of the subject.  Care was taken so that objects would not overlap one another, or straddle two quadrants.  Distribution (grouping) of the objects over these quadrants was varied by sometimes locating all objects within a single quadrant, and at other times distributing the objects over the quadrants as evenly as possible.  Target objects appeared an equal number of times in each quadrant.  Objects were randomly located within quadrants.

_______________________

[1] First, we approximated the counts for the three fine, medium and coarse resolution insetless displays with the following formula: min(disph,hmdh) x min(dispv,hmdv), where hmdh x hmdv is the number of pixels in the HMD screen, and disph x dispv is the number of pixels on the final workstation image for the insetless display. Counts for the remaining six insetted displays were formed with weighted sums of these three counts, with weights corresponding to the percentage of the display area occupied by the insets.

________________

Figure 7 about here

________________

Maximum frame rate was 12, and frame rates did not fall below 11.9.  We enforced this upper limit by introducing a delay into the main display loop whenever a particular frame was too fast. End to end lag (from physical motion to corresponding display motion) in our system without delay averages 197 ms, this was measured with a 30 ms standard deviation and is the minimum lag possible.  With the delay introduced by the maximum frame rate, average lag was 260 ms.

Subjects performed a total of 768 correct trials over all but the insetless, coarse resolution (coarse-0) display. With the latter display, only 16 trials (with no control of correctness) were performed, to confirm that subjects were not performing at better than chance levels. With the other eight displays, trials were divided equally into two sets. A set consisted of eight blocks, each of which corresponded to a single display. Within a display block, subjects performed 48 correct trials. The order in which a set's different display blocks were presented was varied randomly between subjects.

The experiment was run across two or three days. At the beginning of the first experimental session, subjects were told the nature of the experiment. Subjects were randomly assigned to a button condition and given five trials at the beginning of the first and every following session (one object located directly in front of them, seen at fine resolution) to become accustomed to the button assignment. During the first session, the subject was next given 20 practice search trials at the highest device resolution to ensure that they understood the general nature of the task. At the beginning of each 48 trial display block, subjects performed five practice trials with the display. None of these practice trials were included in the analyses. The remaining 48 trials in each the block represented 2 trials for each of the conditions (there were 24 such conditions: 4 locations X 2 grouping X 3 number of objects). Since there were two sets of blocks, subjects performed a total of 4 trials for each of these conditions over the entire experiment. The order of specific trials was randomized for each subject within each block. For any trial on which the subject made an error, that trial was repeated by placing it back into the queue of remaining trials. This in itself was an incentive for accurate performance (subjects required an average of 49.35 trials to complete a block). To further motivate performance, the subject with the best combined accuracy and speed of performance was rewarded with $50.

## 2.2  Results

Initially we performed a display (8) by quadrant (4) by button configuration (2) analysis of variance (ANOVA) on both accuracy and search time to discover if subject button configuration interacted with the other two independent variables.  This analysis revealed no main effect of button configuration, nor any signification interaction of button configuration with display or quadrant.  For all further analyses, this factor was collapsed over.

The data from the visual search task were analyzed by means of 4 four-factor analyses of variance.  We report all significant effects that have a probability level of 0.05 or less.  Bonferroni pair-wise comparisons were used to follow up significant main effects and simple main effects.  The independent variables were display, grouping, number of objects, and target location.  The dependent variables were mean search time for correct trials when the target was present, mean search time for correct trials when target was absent, accuracy when target was present, and accuracy when the target was absent.  We excluded incorrect trials from our time measures because we wished to exclude the false negatives due to accidental button presses.  For all dependent variables, the means are based on collapsing across trials.  The means for these four measures by display are presented in Table 1.  In the search time columns, results labeled with unitalicized letters differed significantly from results labeled with the same, but italicized, letter.  Initially we looked at accuracy of search performance with the insetless, coarse resolution (coarse-0) display.  Since accuracy in this condition was no different from chance, this indicates that subjects were not able to discriminate target from non-target objects with the coarse level of resolution.  Data from this condition were not included in any further analyses.

_________________

Table 1 about here

________________

From this point forward we will use a naming scheme for the individual displays taken from Table 1. The first part of the name refers to the resolution used in the periphery, while the second part refers to the size of the inset in the display. Thus the fine-100 display is the undegraded display, the medium-25 display is the display with medium peripheral resolution and an inset occupying 25% of the display, the coarse-0 display is an insetless display using coarse resolution throughout, and so on.

The 8 X 2 X 3 X 4 analysis of variance on search time when the target was present revealed four significant main effects and five significant interactions. The means for this measure are presented in the column of Table 1 labeled "Search Time Target Present". Figure 8 graphs these results by display type. One of the main effects was for display $[F(7,63) = 17.60, p < .001]$. Follow-up analyses showed that the fine-100, medium-25 and medium-9 displays all had significantly faster search times than the coarse-9 and coarse-2.25 displays. In addition, the medium-2.25, medium-0, coarse-25 and coarse-9 displays had significantly faster search times than the coarse-2.25 display. No other pair-wise differences were significant.

________________

Figure 8 about here

________________

There were also significant main effects of grouping $[F(1,9) = 43.10, p < .001]$, number of objects $[F(2,18) = 40.18, p < .001]$, and location $[F(3,27) = 5.37, p < .05]$. Search times were faster when objects were grouped (mean = 3.422) than when they were ungrouped (mean =

3.741). Search times were faster when there was one object (mean = 3.183), than when there were three objects (mean = 3.610), which were in turn faster than when there were five objects (mean = 3.952). Finally, search time was faster when the target was in one of the quadrants closest to the home position location than when the target was in the two far quadrants.

Of the significant interactions only one involved display. This was an interaction of display and number of objects [$F(6,54) = 4.85$, $p < .001$]. While search time increased with number of objects for all displays, the relative increase was much greater for the coarse-2.25 display.

The 8 X 2 X 3 X 4 analysis of variance on accuracy when the target was present revealed only a significant main effect of number of objects [$F(2,18) = 6.41$, $p < .01$] and a significant interaction of location and grouping [$F(3,27) = 4.54$, $p < .05$]. The means for this measure are presented in the column of Table 1 labeled "% Trials Good Target Present". Analysis of the main effect of number of objects showed that accuracy was higher when there was a single object (mean = 98.4%) than when there were five objects (mean = 95.4%), with the three object condition falling in between these two (mean = 97.3%). Examination of the location and grouping interaction revealed that when objects were grouped, accuracy was lowest when the target was located in the upper far quadrant. When objects were not grouped, accuracy was lowest when the target was in the closest, lower quadrant.

The 8 X 2 X 3 X 4 analysis of variance on search time when the target was absent revealed three significant main effects and three significant interactions. The means for this measure are presented in the column of Table 1 labeled "Search Time Target Absent". There was again a main effect of display [$F(7,63) = 16.77$, $p < .001$]. Follow-up tests showed that all displays had shorter search times than the coarse 2.25 display. In addition, the fine-100 display had significantly shorter search times than the medium-0, coarse-25 and coarse-9 displays. Finally, the medium-75 and medium-9 displays had shorter search times than the coarse-9 display. There were also significant main effects of number of objects [$F(2,18) = 5.51$, $p < .05$] and grouping [$F(1,9) = 105.02$, $p < .001$]. Search times were shorter when there was one object (mean = 5.131) than when there were three objects (mean = 5.918). Search times for five objects fell in between these two (mean = 5.74). Search times were shorter when objects were grouped (mean = 5.146) than when objects were ungrouped (mean = 6.046).

Display and number of objects interacted significantly [$F(14,126) = 10.17$, $p < .001$]. For the coarse-2.25 display, search times increased with number of objects. For all other displays, search times increased between one object and three objects, but search times for five objects fell between those for one and three objects and did not differ from either. There was also a significant interaction of number of objects and grouping [$F(2,18) = 90.16$, $p < .001$]. Search times were faster when objects were grouped and there were three or five objects. Grouping could not take place when there was only one object and this caused the interaction. Finally, there was a significant interaction of location, number of objects and grouping [$F(6,54) = 4.61$, $p < .001$]. The source of this interaction was the shorter search times for five grouped objects when they were located in the upper, near quadrant.

The 8 X 2 X 3 X 4 analysis of variance on accuracy when the target was absent revealed two significant main effects and one significant interaction. The means for this measure appear in the column of Table 1 labeled "% Trials Good Target Absent". Examination of the main effect of number of objects [$F(2,18) = 9.73$, $p < .01$] showed that accuracy was higher when there was one object (mean = 99.1%) or three objects (mean = 97.7%) than when there were five objects (mean = 95.5%). Varying grouping also resulted in a main effect [$F(1,9) = 9.45$, $p < .01$]. Accuracy was higher when objects were grouped (mean = 99.2%) than when objects were ungrouped (mean = 95.7%).  The single significant interaction was between number of objects and grouping [F(2,18) = 9.69, p < .01].  Again, there was higher accuracy for grouped objects when there was more than one object.

In order to determine if there was a significant interaction of inset size and peripheral resolution, we re-analyzed the mean search time data (for both target present and target absent conditions) without the fine-100, medium-0 and coarse-0 conditions.  In these analyses we used a inset size (25%, 9%, 2.25%) by peripheral resolution (medium, coarse) by location (4) by number of targets (3) by grouping (2) analysis of variance.  The analysis on search time for target present condition revealed significant effects of inset size [F(2, 18) = 22.96, p < .001] and peripheral resolution [F(1,9) = 23.12, p < .001].  Search time for the 25% and 9% displays were faster than for the 2.25% displays.  Search times were faster when the peripheral resolution was medium than when it was coarse.  There was also a significant interaction of inset size and peripheral resolution [F(2,18) = 5.24, p < .02].  As can be seen in Figure 8, there was a greater effect of increasing peripheral extent when the peripheral resolution was coarse.  There were also significant effects for the other factors, but as the pattern is the same as in the earlier analyses, we do not report them here.

The same analysis on mean search time for target absent trials also yielded significant effects of inset size [f(2,18) = 10.31, p < .001] and peripheral resolution [F(1,9) = 34.99, p < .001].  The interaction of the two variables was again significant [F(2,18) = 9.67, p < .01). Follow-up analyses revealed once more a greater effect of increasing inset size on search time when the peripheral resolution was coarse.

## 2.3  Discussion

These results clearly show that level of detail in the periphery can be significantly reduced without harming visual search performance. Since reduction in peripheral detail is a technique based upon the perceptual characteristics of the user, and furthermore, since visual search is a task that is particularly sensitive to reduction in peripheral detail, we believe that peripheral detail reduction will be useful across a wide range of tasks. Our manipulations of task difficulty (number of objects, location of objects, grouping of objects, target present/absent) confirm this generalizeability.

___________________

Figure 9 about here

___________________

This is not to suggest that task difficulty has no bearing on use of the peripheral degradation technique. In fact, our results show that as task difficulty increases, detail requirements also increase. The difference between the target present and target absent results is particularly illustrative. When the target object was present, the fine-100 display and the coarse-25 displays were not significantly different. In the more difficult target absent task, the fine-100 and coarse-25 displays were different. Three interactions of the number of objects and display give further evidence of the increased need for detail as task difficulty increases. For each such interaction, performance was poor when the coarse-2.25 display was used in trials with five objects.

Accuracy was uniformly high among all displays (excepting coarse-0). Addition of detail only to the central area also allowed subjects to maintain fast times. Search times with the fine-100 and the insetted medium resolution displays were not different, while search times with the fine-100 and medium-0 displays in the target absent condition did differ. This suggests that the time required to make a visual discrimination is related to the ease with which that discrimination can be made.

While the difference between the fine and medium levels of resolution was not large (21.74 vs. 23.62 arcmins), this difference still resulted in significantly different levels of performance for the fine-100 (28912 pixels) and medium-0 (26688 pixels) displays in the target absent condition. The addition of even a small fine resolution inset eliminated this difference. The small distinction between the fine and medium levels of resolution thus provides further support for the paramount importance of detail in the center of the display.

Two interactions of degraded peripheral extent and degraded peripheral LOD showed that the effect of varying degraded extent increased as peripheral LOD decreased. Clearly peripheral LOD should be considered when sizing the area of high detail. In order to maintain performance, the size of the high detail inset must increase as the level of peripheral detail is decreased.

In summary, these results suggest that reducing LOD in display peripheries is an effective way of achieving the conflicting goals of improving computational speed and maintaining high user performance. Peripheral LOD and degraded peripheral extent can be adjusted to allow users to achieve required performance levels. The results also suggest that the area of high detail required to maintain both accuracy and speed in this task is quite small.

## 3. EXPERIMENT 2

Psychophysical research has shown that color vision is worse in the periphery of the retina by many different measures [Zrenner, Abramov, Akita, Cowey, Livingstone & Valberg, 1990]. In this experiment we decreased level of detail by using only gray scale values in the periphery of the display.  As in the first experiment, we compared visual search task performance with an undegraded display, an evenly degraded display, and peripherally degraded displays.  The key differences were in the type of peripheral degradation and the discrimination component of the task.  In this experiment, subjects searched for a target which differed in color only.

## 3.1  Experimental Methodology

3.1.1  *Participants*.  The participants in the experiment were 10 graduate students.  All the subjects had 20/40 vision, uncorrected or corrected with contact lens.

3.1.2  *Design*.  This study utilized a six factor, mixed design, very similar to the design of the first experiment.  The primary independent variables were again all within-subjects variables.  These variables were display (varying in color content and degraded display area, see Table 2), number of objects, and grouping.  The three control variables were trial condition, number of trials, and button assignment.

3.1.3  *Apparatus*.  The apparatus used in this experiment was the same as that in Experiment 1 except for two changes.  First, while this experiment's objects had the same relative size as the first experiment's, they were not textured with an image of a face.  Instead the objects varied only in 24-bit color.  Subjects were asked to find an orange target box in a field of blue distracter boxes.  In 8-bit gray scale display regions, this translated into finding a brighter target in a field of less bright objects.

Second, in this experiment a different head-mounted display was used.  Subjects wore a Virtual Research VR 4 HMD.  The VR 4 mounts two color LCD displays on the user's head, each with vertical field of view of 36 degrees, and a horizontal FOV of 48 degrees.  These two FOVs overlap fully.  Because this HMD does not include LEEP optics, it does not introduce distortion.

Each LCD contains an array of 247 x 230 color triads, with a resolution of 11.66 arcmin per pixel. The fine level of resolution from the first experiment was used for every one of this experiment's displays. Since the final workstation image for fine resolution contained more pixels than the VR 4 display, there was again some resolution loss. However, the HMD pixel count in this experiment was the same for every display, and thus the counts of workstation image color bytes shown in the column of Table 2 labeled "Total Bytes of Color in Wkstn" are proportionally accurate. We have graphed these workstation color byte counts in Figure 10.

__________________

Figure 10 about here

__________________

We used the VR 4 in a biocular mode by sending the same image to each of the video inputs. The search space and the objects in it were scaled down in proportion to the reduced FOV of the VR 4. Relative sizes of the objects in the virtual environment remained the same.

3.1.4 *Procedures*. Subjects performed a total of 480 correct trials, divided equally into two sessions, constrained and randomized as before. Each session consisted of five blocks of 48 trials. Subjects either completed both sessions in one day, with a break between the sessions, or one session per day, over two days.

## 3.2 Results

The data from the visual search task were analyzed by means of 4 three-factor analyses of variance. Bonferroni pair-wise comparisons were used to follow up significant main effects and simple main effects. Significant effects are reported when the probability level is 0.05 or less. The independent variables were display, number of objects, and grouping. The dependent variables were mean search time when the target was present, mean search time when target was absent, accuracy when target was present, and accuracy when the target was absent. For all dependent variables, the means are based on collapsing across trials and button assignment, as an initial analysis using button assignment as an independent variable failed to reveal significant effects of button assignment. The means for these four measures are presented by display in Table 2.

_____________________

Table 2 about here

_____________________

From this point forward we will use a naming scheme for the individual displays taken from Table 2. The first part of the name refers to the color used in the periphery, while the second part refers to the size of the inset in the display. Thus the color-100 display is the undegraded display, the gray-25 display is the display with a gray periphery and an inset occupying 25% of the display, the gray-0 display is an insetless display using gray throughout, and so on.

The 5 X 3 X 2 analysis of variance on search time when the target was present revealed two significant main effects and two significant interactions. The means for this measure appear in the column of Table 2 labeled "Search Time Target Present", and are graphed in Figure 11. There was a significant effect of display [$F(4,36) = 29.20$, $p < .001$]. The gray-0 display yielded longer search times than the other four displays. In addition, the color-100 display had faster search times than the gray-2.25 display. No other displays differed significantly. There was also a significant effect of grouping [$F(1,9) = 42.82$, $p < .001$]. Search times were faster when the objects were grouped in a single quadrant (mean = 3.011) than when they were placed in different quadrants (mean = 3.253).

_______________

Figure 11 about here

_______________

One of the significant interactions was between grouping and number of objects [F(2,18) = 7.17, p < .01]. Follow-up testing showed that when the objects were ungrouped, search time increased with the number of objects. When the objects were grouped, search time decreased when the number of objects increased. There was also a significant three-way interaction [F(8,72) = 2.12, p <.05]. Follow-up analyses suggest that the above two-way interaction is due to the three insetted displays. In the other two displays this relationship does not hold.

The 5 X 3 X 2 analysis of variance on accuracy when the target was present yielded two significant main effects and one significant interaction. The means for this measure are in the column of Table 2 labeled "% Trials Good Target Present". Display again had a significant effect [F(4,36) = 26.96, p < .001]. Subjects had lower accuracies in the gray-0 than in the other four displays. The other four displays did not differ in accuracy. Accuracy in the gray-0 condition was midway between chance and the other four conditions (mean = 74.8%). This suggests that subjects could discriminate the target object from other objects based on luminance differences, but not as easily as in displays using color. Analysis of the significant effect of grouping [F(1,9) = 7.98, p < .05] revealed that the accuracy was higher when objects were grouped (mean = 93.6%) than when not grouped (90.3%). Grouping also interacted significantly with number of objects [F(2,18) = 4.09, p < .05]. In the ungrouped conditions, accuracy was highest in the single target condition, while when objects were grouped, accuracy was highest when there were three objects.

The 5 X 3 X 2 analysis of variance on search time when the target was absent yielded three significant main effects and four significant interactions. The means for this measure are in the column of Table 2 labeled "Search Time Target Absent". Examination of the significant effect of display [$F(4,36) = 28.78$, $p < .001$] showed that search time was longer in the gray-0 display than all other displays. The color-100 display had faster search times than the gray-9 and gray-2.25 displays. Search times with the gray-25 display were faster than the search times with the gray-2.25 display.

Number of objects also had a significant effect [$F(2,18) = 4.38$, $p < .05$]. Follow-up analysis revealed that search times were less when there were three objects than when there was one or five objects. Examination of the final significant effect of grouping [$F(1,9) = 53.28$), $p < .001$] showed that search times were less when the objects were grouped (mean = 4.754) than when ungrouped (mean = 5.602). Display and number of objects interacted significantly [$F(8,72) = 8.15$, $p < .001$]. While the relative ordering of search times for the five displays was the same across one, three and five objects, the absolute difference in search times between the displays increased with the number of objects. Display also interacted with grouping [$F(4,36) = 4.22$, $p < .01$]. Again, while the ranking of the displays did not differ for grouped and ungrouped conditions, the relative differences in search time between displays were larger when objects were ungrouped.

Follow-up analyses of the significant interaction between number of objects and grouping [$F(2,18) = 48.49$, $p < .001$] revealed that in the ungrouped condition, search times increased with the number of objects. In the grouped condition, search times were less when there were three objects. There was also a significant three-way interaction [F(8,72) = 4.43, p < .001]. The locus of this interaction seemed to reside in the differential effects that number of objects and grouping had with the gray-0 and gray-2.25 displays. When objects were ungrouped, increasing the number of objects increased search times for these two displays. For the other displays, increasing the number of objects had a negligible effect on search times.

The 5 X 3 X 2 analysis of variance on accuracy when the target was absent revealed three significant main effects and two significant interactions. The means for this measure are in the column of Table 2 labeled "% Trials Good Target Absent". Display once again had a significant effect [F(4,36) = 26.13, p < .001]. Accuracy with the gray-0 display was lower than with the other four displays. No other displays differed in accuracy. Number of objects also had a significant effect [F(2,18) = 18.02, p < .001]. Accuracy was lower when there were five objects (mean = 84.7%) than when there was only one object (mean = 95.3%). When there were three objects accuracy was in between these two conditions (mean = 91.8%). Analysis of the main effect of grouping [F(1,9) = 6.88, p < .05] revealed that accuracy was lower when objects were ungrouped (mean = 88.8%) than when objects were grouped (mean = 92.4%). There was a significant interaction of display by number of objects [F(8,72) = 5.36, p < .001]. With the gray-0 display, increasing the number of objects from one to three significantly lowered accuracy. For the other four displays there was no such effect. Finally, there was a significant interaction of number of objects and grouping [F(2,18) = 3.68, p < .05]. Because grouping cannot occur with a single object, there was no effect in the single object condition.

## 3.4 Discussion

Overall, the pattern of this experiment's results strongly suggest that search performance is not greatly affected by the use of peripheral degradation as compared to full color. Accuracy was lower with the grey-0 display than with all other displays, but adding even the smallest amount of central high detail eliminated any differences in accuracy between the color-100 and peripherally degraded displays. As the area of the high detail inset was increased even further, speed also improved. For the target absent trials, search time for the color-100 display was faster than for the grey-9 and grey-2.25 degraded displays. For the target present trials, search time for the full color display was only faster than for the grey-2.25 degraded display. However, in no case was there a significant difference in performance between the color-100 display and the grey-25 display.

_________________

Figure 12 about here

_________________

As in the first experiment, several interactions between display and grouping or number of objects suggested that task complexity is an important factor when sizing the peripherally degraded display area. These results, when combined with the differences in the target absent and target present conditions suggest that visual search task difficulty interacts with the level of detail. When the search task is easier, the amount of detail that can be eliminated without hurting performance is greater.

## 4. CONCLUSIONS

The purpose of this work was to investigate a way to circumvent the tradeoffs that exists between the extent of perceptual fidelity, system responsiveness, and performance. In virtual environments, system responsiveness is determined primarily by the level of visual complexity, with decreased responsiveness when there are higher levels of visual complexity. As the system becomes less responsive, it becomes more and more difficult for people to interact effectively with the system and performance drops. However, we also know that user performance decreases with reduced visual complexity. Designers of virtual environments must therefore balance the demands for increased system responsiveness with the need for more visual complexity in order to produce a system that yields optimal user performance. One approach that has been suggested to balance these tradeoffs is to reduce the level of visual detail in the periphery (e.g., Funkhouser & Sequin; 1993; Watson et al., 1995). In this study we have clearly shown that this technique holds great promise.

The results of both experiments show that we can reduce visual complexity in the periphery without adversely affecting visual search task performance. We would argue that these results provide a strong test of the effects of LOD management by reducing peripheral complexity on visual search performance as in both experiments we reduced detail that was critical to visual search performance. In some task conditions, visual complexity was reduced by almost half (as measured by number of pixels and bytes of color) without lowering accuracy or speed of visual search.

This is not to say that peripheral information is not needed for performance of our tasks. In the experiments there was a point at which reducing peripheral information hurt performance. In general, speed was affected before accuracy.

There are several characteristics of our results that merit further examination. First, while our search task was certainly a valid test of peripheral degradation, there may be other sorts of tasks that are more severely affected by this method of LOD management. Research is needed into the effect of peripheral degradation on other sorts of basic tasks, such as tracking, grasping, and travel. Our results also indicate that an analysis of task and environment complexity should play an important role in the configuration of a peripherally degraded display. Isolation of the parameters and dimensions of this complexity should make this sort of analysis more simple. We found significant interactions between the size of the degraded periphery and the LOD in that periphery. Could there be an LOD threshold below which this interaction does not exist? More focused follow-up research into this question might allow the definition of more concrete guidelines for VE system designers using peripheral LOD degradation.

Nevertheless, at this point we can make some suggestions about the configuration of a peripherally degraded display. In order to maintain both accuracy and speed, VE designers might run pilot studies with evenly degraded displays in order to isolate the LOD at which accuracy remains high, but speed begins to decline significantly. Once this LOD is found, a high detail inset of 25% of display area (or perhaps 25 horizontal degrees of visual angle) might be added to enable users to maintain speed. In our experiments, this sort of display never differed significantly from an undegraded display.

Of course, even this suggestion begs many additional questions. For example, how meaningful is percentage of display area as compared to degrees of visual angle when discussing degraded display area and user performance? In our studies, although we did not use eye tracking, peripheral degradation was still effective. Would peripheral degradation with eye tracking be more effective and allow significant additional computational savings? We also used only two levels of detail. Would a more continuous degradation improve the interface? Finally, our studies only varied LOD in image and color space. How effective would peripheral degradation be with texture or model-based LOD?

Although many questions remain unanswered, our studies show that peripheral level of detail degradation should prove to be an effective way of raising frame rates and reducing latencies.

Systems that use this management technique will be able to reduce computation with minimal effect on user performance.

## ACKNOWLEDGMENTS

We would like to express our appreciation to the members of the VE group at Georgia Tech's GVU Center.

**Table 1:** Average search times and accuracies for the 9 displays examined in the first experiment. Search times labeled with unitalicized letters differed significantly from search times labeled with the same, but italicized letter.

| Display Resolution | | Pctg Of Display For Inset | Horiz Degrees For Inset | Total Pixels in Wkstn | Approx Pixels in HMD | Search Time Target | | % Trials Good Target | |
| Inset | Periph | | | | | Present | Absent | Present | Absent |
|---|---|---|---|---|---|---|---|---|---|
| Fine | | 100 | 75.3 | 76800 | 28912 | 3.168 a | 4.932 a | 96.2 | 96.3 |
| Fine | Medium | 25 | 37.65 | 45216 | 27244 | 3.267 a | 5.151 b | 96.9 | 97.8 |
| Fine | Medium | 9 | 22.59 | 37448 | 26888 | 3.340 a | 5.092 b | 96.6 | 97.6 |
| Fine | Medium | 2.25 | 11.29 | 31358 | 26738 | 3.551 b | 5.431 c | 98.2 | 97.2 |
| | Medium | 0 | 0 | 27648 | 26688 | 3.564 b | 5.671 *ac* | 96.4 | 97.0 |
| Fine | Coarse | 25 | 37.65 | 26784 | 9532 | 3.586 b | 5.776 *ac* | 97.5 | 98.0 |
| Fine | Coarse | 9 | 22.59 | 15084 | 5398 | 3.787 *ab* | 5.850 *abc* | 97.5 | 96.9 |
| Fine | Coarse | 2.25 | 11.29 | 7334 | 3654 | 4.388 *ab* | 6.867 *abc* | 97.0 | 97.4 |
| | Coarse | 0 | 0 | 3072 | 3072 | 5.429 | 5.909 | 50.6 | 54.2 |

**Table 2:** Average search times and accuracies for the five displays of the second experiment. Search times labeled with unitalicized letters differed significantly from search times labeled with the same, but italicized letter.

| Display Color | | Pctg of Display For Inset | Horiz Degrees For Inset | Total Bytes of Color in Wkstn | Search Time Target | | % Trials Good Target | |
|---|---|---|---|---|---|---|---|---|
| **Inset** | **Periph** | | | | **Present** | **Absent** | **Present** | **Absent** |
| Color | | 100 | 48 | 230400 | 2.141 a | 2.714 a | 95.1 a | 93.1 a |
| Color | Gray | 25 | 24 | 131040 | 2.528 b | 4.533 b | 95.8 a | 97.0 a |
| Color | Gray | 9 | 14.4 | 100164 | 2.934 b | 5.217 *ac* | 96.7 a | 95.6 a |
| Color | Gray | 2.25 | 7.2 | 88068 | 3.115 *ab* | 5.633 *abc* | 95.4 a | 94.4 a |
| | Gray | 0 | 0 | 76800 | 4.941 *ab* | 6.793 *abc* | 76.8 *a* | 72.9 *a* |

**Figure Captions**

**Figure 1:** Illustration of the technique used to generate images with high detail insets. Two images are generated on the screen. One is textured into the periphery, one into the inset. The boundary is blended with transparency.

**Figure 2:** Experimental environment as seen with the coarse-75 display, grouped target present condition. The face with the mouth closed is the target object.

**Figure 3:** The complete image generation process, from initial display on the workstation screen to optical imaging by the HMD optics. Dashed outlines indicate images of variable size.

**Figure 4:** Top down schematic view of the experimental environment. Shown are the platform, the user, the home object, the search quadrants, and three example objects. The platform and user are not visible in the virtual environment.

**Figure 5:** The three levels of the inset size variable. Center box contains high detail inset.

**Figure 6:** Detail in estimated HMD pixels for the nine displays in experiment 1.

**Figure 7:** The four quadrants used to control object location.

**Figure 8:** Search time in seconds for the nine displays in experiment 1 when the target was present.

**Figure 9:** Search time versus estimated pixels displayed in the HMD for the nine displays in experiment 1 when the target was present.

**Figure 10:** Detail in bytes of color in the final workstation image for the five displays experiment 2.

**Figure 11:** Search time in seconds for the five displays in experiment 2 when the target was present.

**Figure 12:** Search time versus color bytes in the workstation image for the five displays in experiment 2 when the target was present.

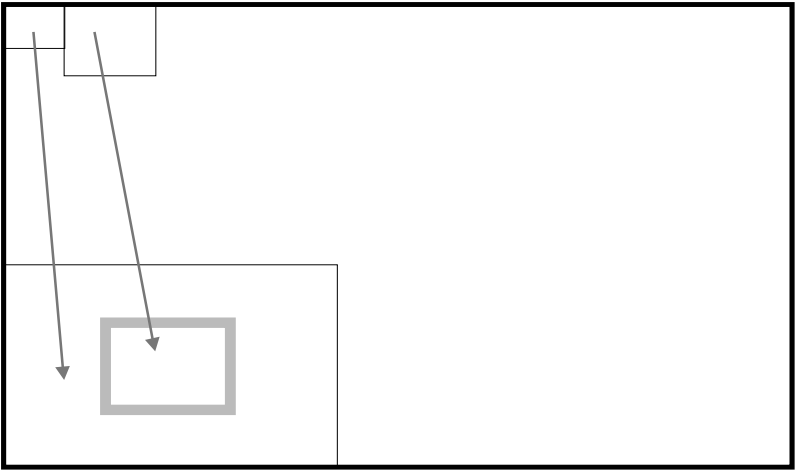

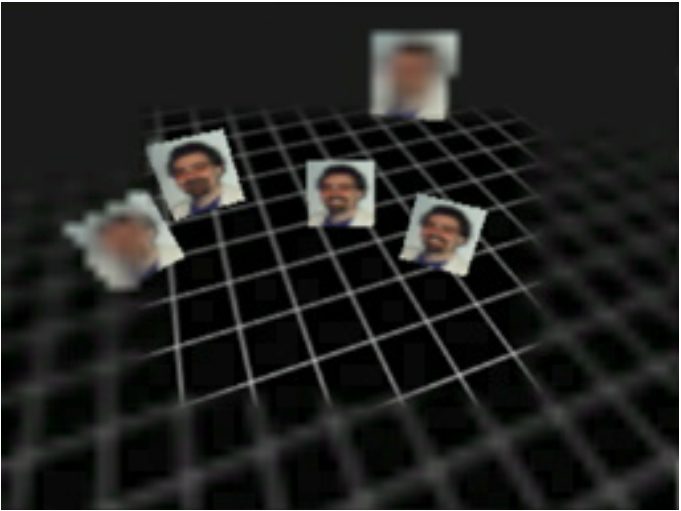

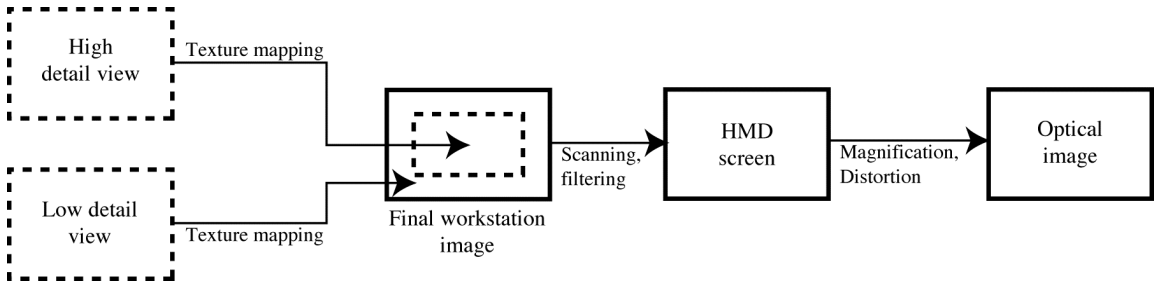

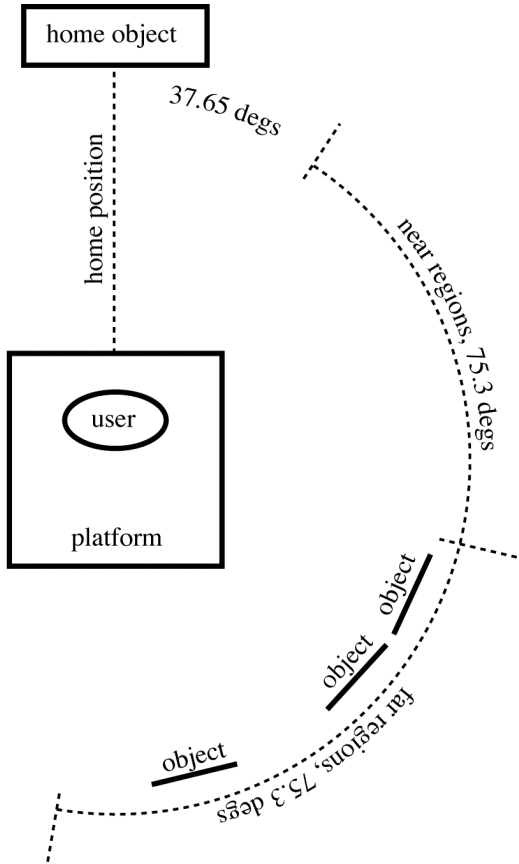

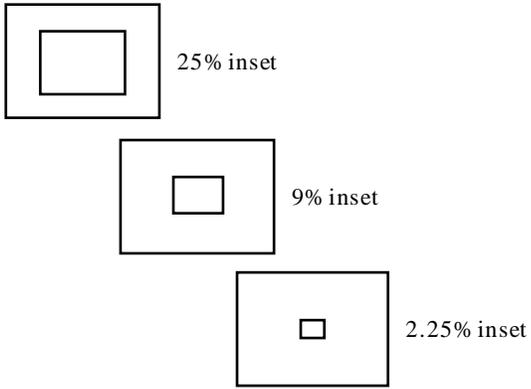

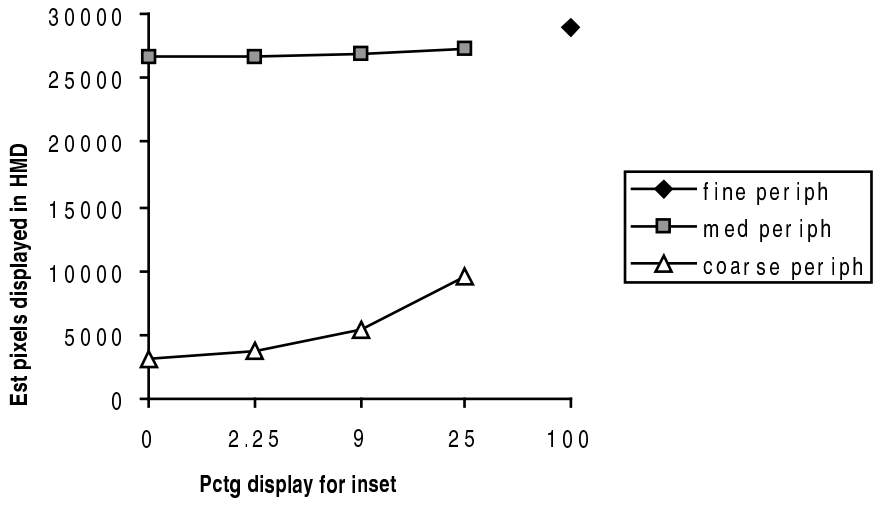

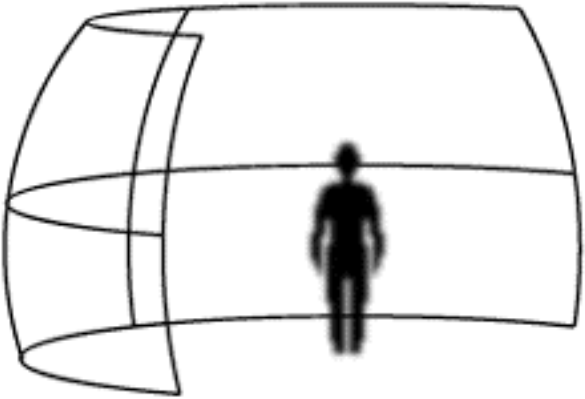

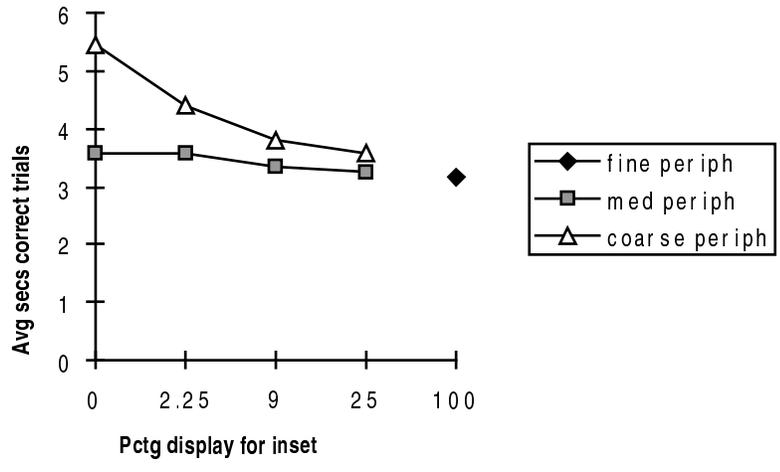

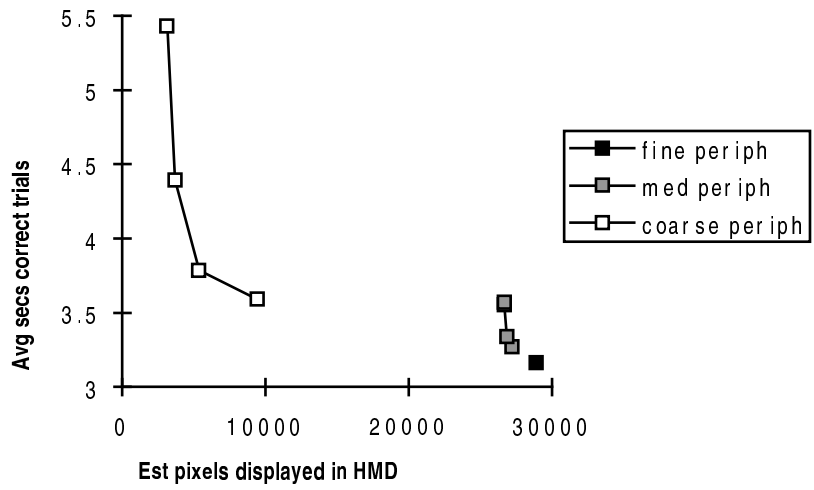

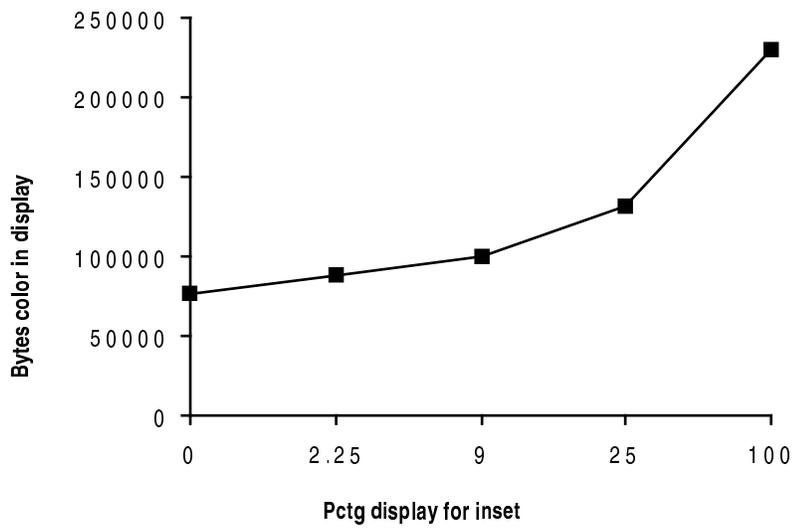

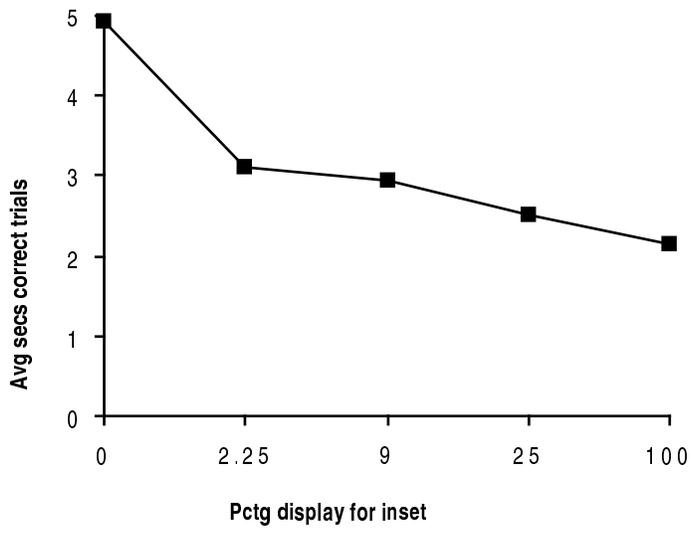

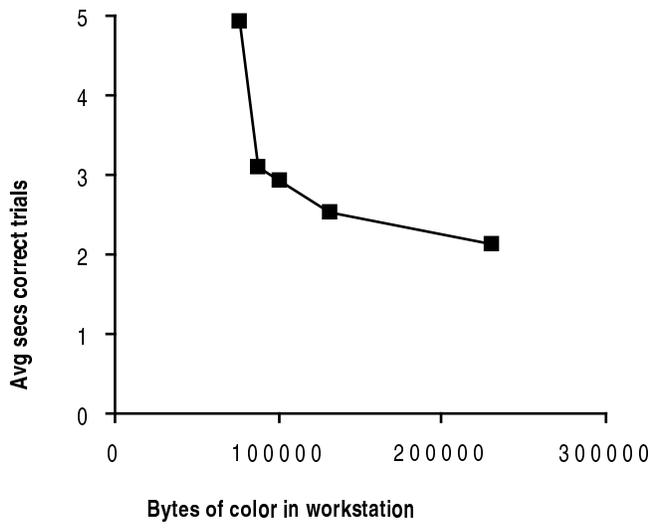